\documentclass[%
 reprint,
 amsmath,amssymb,
 aps,
prb,
]{revtex4-2}

\usepackage{graphicx}
\usepackage{dcolumn}
\usepackage{bm}

\begin{document}


\title{$^3$He adsorbed on molecular hydrogen surfaces }

\author{M. C. Gordillo$^{1,2}$}
\email[]{Corresponding author: cgorbar@upo.es}
\affiliation{$^{1}$Departamento de Sistemas F\'{\i}sicos, Qu\'{\i}micos
y Naturales, Universidad Pablo de
Olavide, Carretera de Utrera km 1, E-41013 Sevilla, Spain}
\affiliation{$^2$Instituto Carlos I de Física Teórica y Computacional, Universidad de Granada, E-18071 Granada, Spain}

\author{J. Boronat$^3$}
\affiliation{$^3$Departament de F\'{\i}sica,
Universitat Polit\`ecnica de Catalunya,
Campus Nord B4-B5, 08034 Barcelona, Spain}

\date{\today}

\begin{abstract}
Using a diffusion Monte Carlo (DMC) technique, we calculated the phase diagram of $^3$He adsorbed on a first 
solid layer of a molecular hydrogen isotope (H$_2$,HD and D$_2$) on top of graphite.  The results are qualitatively
similar in all cases: a two-dimensional gas spanning from the infinite dilution limit to a second-layer helium density of 0.048 $\pm$ 0.004 \AA$^{-2}$.   
That gas is in equilibrium with a 7/12 commensurate structure, more stable than any 
incommensurate triangular solid of similar density.  These findings are in 
reasonably good agreement with available experimental data.  
 \end{abstract}

\maketitle
               
\section{Introduction}  

At the heart of any Monte Carlo calculation lies the same principle that allow 
us to compute a simple integral by a hit-and-miss method, modified by the 
introduction of techniques to reduce the statistical variance. 
The first relevant application of that proposal 
was the calculation of the properties of a gas of hard spheres in the seminal 
paper by Metropolis {\em et al} \cite{metropolis}.  Thus, the basic idea behind 
the Monte Carlo method   consists in transforming the expression that describe 
the phenomena we are 
interested in into an integral,  and apply a modification of that simple 
integration recipe to calculate the desired magnitude \cite{allen}. 
For classical averages this is straightforward,  since the equations that define 
them are already integrals defined in a multidimensional space.  The 
application to quantum systems is more involved but it is now currently used in 
a set of methods known globally as quantum Monte Carlo (QMC).

The simplest QMC method consists in the application of the variational 
principle of Quantum Mechanics to calculate the energy and other observables 
for a proposed many-body wavefunction \cite{hammond}. The accuracy of 
this method, known as variational Monte Carlo (VMC), depends 
on the quality of that wavefunction.
This constraint can be removed by using the diffusion Monte Carlo 
(DMC) algorithm \cite{kalos}, which tackles directly the imaginary-time 
Schr\"odinger equation by using the connection between a random walk and the 
diffusion equation. Therefore, DMC
is able, at least in principle, to produce 
the real ground state of any many-body system by improving upon that initial 
guess. That technique has been applied successfully to many bosonic and 
fermionic ensembles of particles \cite{boro94,hammond}.  In this work, we are 
going to use the DMC algorithm to study the 
properties of a system that includes either fermions ($^3$He atoms, HD
molecules) or a a mixture of bosons and fermions ($^3$He adsorbed on top
of a H$_2$ or a D$_2$ layer). 

The calculations presented in this work are prompted by experimental results on $^3$He adsorbed on a double HD layer on top of graphite, both using NMR techniques \cite{casey2,ikegami1,ikegami2,masutomi} or calorimetric measurements \cite{casey2,casey1,casey,fukuyama2,fukuyama3}. Those findings are part of a long series of studies on $^3$He on clean and preplated graphite, both from the experimental 
\cite{PhysRevB.18.285,PhysRevB.41.1842,PhysRevLett.65.64,PhysRevB.49.12377,PhysRevB.53.2658, PhysRevLett.76.1884,PhysRevLett.86.2447,PhysRevLett.109.235306,fukuyama,Todoshchenko}
and theoretical \cite{preplated,secondhe3} points of view.  All those experimental studies of $^3$He on HD preplated  graphite agree in finding both a liquid/gas and a low-density commensurate solid.  In this work, we will address the nature of both phases by means of DMC calculations. 

The rest of the paper is organized as follows. In the next section, we 
 describe the DMC algorithm and its application to obtain the 
equation of state of $^3$He on a single molecular hydrogen layer on top of 
graphite. In Sec. III, we report the results obtained for the equations of 
state of the $^3$He films adsorbed on different molecular hydrogen isotopes. 
Finally, Sec. IV comprises a brief summary and a discussion of the main 
conclusions.

\section{Method}

The most ambitious approach to the quantum many-body problem, from a 
microscopic point of view, is to solve its corresponding 
Schr\"odinger equation.  The diffusion Monte Carlo (DMC) method does so 
stochastically,  starting by its imaginary-time counterpart \cite{boro94}: 
\begin{equation}
-\frac{\partial \Psi(\bm{R},t)}{\partial t} = (H-E) \Psi(\bm{R},t) \,,
\label{eq:Sch1}
\end{equation}
with $\bm{R}$ standing for the positions of all the atoms/molecules in the 
system. The Hamiltonian $H$ of the system, composed by $^3$He atoms and 
hydrogen molecules,  is given by
\begin{eqnarray} \label{hamiltonian}
H = \sum_{\alpha} \sum_{i=1}^{N_{\alpha}}  \left[ -\frac{\hbar^2}{2m_{\alpha}} \nabla_i^2 +
V_{{\rm ext}}^{(\alpha)} (x_i,y_i,z_i) \right] + \\ \nonumber
 \sum_{\alpha} \sum_{i<j}^{N_{\alpha}}
  V_{{\rm pair}}^{(\alpha,\alpha)} (r_{ij}) + \sum_{\alpha} \sum_{\beta} 
\sum_i^{N_{\alpha}} 
\sum_j^{N_{\beta}}  V_{{\rm pair}}^{(\alpha,\beta)} (r_{ij}) .
\end{eqnarray}

As in previous literature, we considered graphite as a rigid structure, i.e., 
its influence on the behavior of $^3$He and hydrogen molecules will be modeled 
by an external potential, $V_{{\rm ext}}^{(\alpha)}(x_i,y_i,z_i)$, different 
for each species $\alpha$ = [$^3$He,(H$_2$,HD,D$_2$)].  To release 
that constraint by allowing
the carbon atoms to move around their crystallographic positions does not change 
the behavior of the first layer of molecular hydrogen at the densities considered in this work \cite{PhysRevB.88.041406}.  In 
Eq. \ref{hamiltonian},  the coordinates 
$x_i$, $y_i$, and $z_i$ correspond to each of the $N_{\alpha}$ or $N_{\beta}$ (Helium {\em 
or} Hydrogen) adsorbate particles with mass $m_{\alpha}$.  All the individual 
adsorbate-carbon interactions were explicitly considered, in a full rendition 
of graphite as a corrugated structure made up of parallel layers separated 3.35 
\AA$ $ in the $z$ direction.  In all cases, those $V_{{\rm 
ext}}^{(\alpha)}(x_i,y_i,z_i)$ potentials were taken to be of 
the Lennard-Jones type and no distinction was made between different hydrogen 
isotopes \cite{potentialhd1}.  In particular, the He-C interaction was taken 
from Ref.~\onlinecite{carlos}, while the (H$_2$,HD,D$_2$)-C potential was the 
one derived  in Ref.~\onlinecite{coleh2}. 

In the Hamiltonian (\ref{hamiltonian}),  we have as many expressions for 
$V_{\rm{pair}}^{(\alpha,\beta)}$ as possible 
adsorbate pairs, i.e., He-He,  He-(H$_2$,HD,D$_2$) and  (H$_2$,HD,D$_2$)-(H$_2$,HD,D$_2$).  
For the $^3$He-$^3$He interaction,  we used the standard Aziz 
potential \cite{aziz},  while for any  hydrogen-hydrogen interaction we resort 
to the Silvera and Goldman expression \cite{potentialhd1,silvera}.   The 
Helium-H$_2$ potential was taken from Ref.  \cite{whaley}, previously used in 
the study of small clusters including $^4$He-H$_2$ mixtures \cite{gordillo99}.  
What all those potentials have in common is that they are isotropic 
interactions, depending only on  the distance $r_{ij}$ between particles $i$ 
and $j$.  In the case of molecular hydrogen, an elipsoid,  the Silvera and Goldman potential 
was built to reproduce the isotropic properties of solid phases and does
so very successfully.  
 In all cases, the hydrogen molecules were not kept fixed but allowed to move 
around their crystallographic positions. 

The solution of Eq. \eqref{eq:Sch1} can be formally written as
\begin{equation} \label{pro}
\Psi(\bm{R}', t + \Delta t)= \int G(\bm{R}',\bm{R},\Delta t)  \Psi (\bm{R}, t) 
d \bm{R} \ ,
\end{equation}
with $t$ the imaginary time. The Green's function is given by
\begin{equation}
      G(\bm{R}',\bm{R},\Delta t)  = \langle  \bm{R}' |   \exp[-(H -E)\Delta t] 
| \bm{R} \rangle \ ,
\label{Green}
\end{equation}
with $E$ an energy close to the ground-state value.
By remembering that $\Psi (\bm{R},t)$ can be expanded in terms of a complete set
of the Hamiltonian's eigenfunctions,  $\Phi_i (\bm{R})$,  with eigenvalues 
$E_i$,   as
\begin{equation}
\Psi (\bm{R},t) = \sum_i c_i e^{-(E_i-E) t} \, \Phi_i (\bm{R}), 
\end{equation}
we can see that, successive applications of Eq. \ref{pro} on any initial 
approximation to the exact wavefunction, will project to the ground state in the 
$t$ $\rightarrow \infty$ limit, i.e,  this method produces a zero-temperature 
estimation.  
However, given the very low 
temperatures at which the relevant experiments are performed,  usually of the 
order of the mK,  that solution is expected to be a very good approximation to 
what is observed.    Any iterative application of Eq. \ref{pro} constitutes a Monte Carlo step in 
the DMC algorithm. 

Unfortunately,  this procedure, even though it is formally correct, produces  
very noisy estimations \cite{hammond}.  To reduce the statistical variance of 
the results to a manageable level,  one introduces importance sampling.   This 
is done by means of a time-independent trial wave function, $\psi(\bm{R})$,  as 
close as possible to the exact solution of Eq. \ref{eq:Sch1}.  
We define then an auxiliary function, $f(\bm{R},t)$,  as
\begin{equation}
f(\bm{R},t) =  \psi(\bm{R}) \Psi (\bm{R}, t), 
\end{equation}
that introduced in Eq. \ref{eq:Sch1} gives
\begin{equation} \label{eqf}
-\frac{\partial f(\bm{R},t)}{\partial t} = A(\bm{R},t) f(\bm{R},t) \,,
\end{equation}
 with 
\begin{eqnarray} \label{eqf2}
 A(\bm{R},t) = -\frac{\hbar^2}{2m_i} \nabla_i^2  + 
  \frac{\hbar^2}{2 m_i} F(\bm{R}) + \\ \nonumber 
 [E_L(\bm{R})-E]  \ .
\end{eqnarray}
 At difference with Eq.  (\ref{eq:Sch1}), Eq.  \ref{eqf}
 includes a drift term, with  $F(\bm{R}) = 2 \psi(\bm{R})^{-1} \nabla 
\psi(\bm{R})$, that  guides the stochastic process to the regions where the 
trial function is larger.  $E_L(\bm{R}) = \psi(\bm{R})^{-1} H \psi(\bm{R})$ is 
the so-called local energy, whose mean value is the exact energy of the 
system. 

We considered $^3$He adsorbed on a single molecular hydrogen 
layer,  contrarily to the experimental setups of Refs.  \onlinecite{
casey2,ikegami1,ikegami2,masutomi,casey1,casey,fukuyama2,fukuyama3}, 
in which Helium is adsorbed on top of two or more \cite{masutomi} HD sheets.  
To include two layers would have doubled the number of hydrogen molecules in our simulations (see below),  
and  would have implied a considerable increase in the computational complexity and in the simulation time.   Moreover, 
the vertical distance between the $^3$He sheet and a second hydrogen layer closest to the graphite would 
have been large enough (around 6 \AA, see for instance the distribution of H$_2$ layers in Ref. \onlinecite{prb2022})  to make the influence of that layer on the Helium equation of state negligible,  apart from a nearly constant correction in the value of the binding energy.  That correction is expected to be small given the shallowness of the  He-(H$_2$,HD,D$_2$) potential \cite{whaley}.   In any case,  ours is
a new quasi-2D $^3$He system whose behaviour could be directly compared with an 
experimental 
setup with a single hydrogen layer on top of graphite. 


Taking all that in mind, and 
following  previous work on $^3$He films   
\cite{preplated,secondhe3}, we considered a two-layer trial wave function of the form
 \begin{eqnarray}
\psi({\bf r}_1, {\bf r}_2, \ldots, {\bf r}_N) = \psi_1({\bf r}_1, {\bf
r}_2, \ldots, {\bf r}_{N_1}) \times \nonumber  \\
\psi_2({\bf r}_{N_1+1}, {\bf r}_N, \ldots, {\bf r}_{N}) \ ,
\label{trialtot}
\end{eqnarray}
with $N_1$ the number of hydrogen molecules in the single layer adsorbed on  the 
graphite surface 
and $N$  the total number of 
particles (H$_2$/HD/D$_2$ molecules and $^3$He atoms).  The 
number of $^3$He atoms in the second layer is thus 
$N_2$ = $N$-$N_1$. 
The trial wave function for the upper $^3$He layer is \cite{preplated} 
\begin{eqnarray}
\psi_2({\bf r}_{N_1+1}, {\bf r}_{N_1+2}, \ldots, {\bf r}_{N}) = 
D^{\uparrow} D^{\downarrow}  \prod_{i=N_1+1}^{N}  u_3({\bf r}_i)  \times \nonumber \\
\prod_{i<j}^{N_2} \exp \left[-\frac{1}{2} 
\left(\frac{b_3}{r_{ij}} \right)^5 \right], 
\label{trial3}
\end{eqnarray}
where  $D^{\uparrow}$ and $D^{\downarrow}$ are Slater determinants 
including
two-dimensional plane waves depending on the
second layer particle coordinates (with spins up and down) and whose periodicity is determined by 
the size of the simulation cell.  In all cases, we considered the
same number of spin-up and spin-down $^3$He atoms.  The coordinates in the
Slater determinants were corrected by backflow terms in the standard way \cite{backflow,borobook},
\begin{eqnarray}
\tilde x_i  & = & x_i + \lambda \sum_{j \ne i} \exp [-(r_{ij} -
r_b)^2/\omega^2] \, (x_i - x_j) \\
\tilde y_i  & = & y_i + \lambda \sum_{j \ne i} \exp [-(r_{ij} - r_b)^2/\omega^2] 
\, (y_i - y_j).\end{eqnarray} 
The optimal values for the parameters in the backflow term 
were those of the bulk three-dimensional
system~\cite{casulleras,prl2016}, i.e., $\lambda = 0.35$; $\omega =
1.38$ \AA,  and $r_b = 1.89$ \AA. 

The one-body function $u_3({\bf r})$ is the numerical 
solution of the  
Schr\"odinger equation that describes a single $^3$He atom on top of a
hydrogen first layer of density 0.095 \AA$^{-2}$. 
This is the largest experimental HD density before a promotion 
to a second HD layer on top of graphite happens \cite{experiment}. This 
density is the same as for H$_2$ promotion to the a second layer 
\cite{prbsecondh2,prb2022,wiechert},
and comparable to the density D$_2$ needs to jump to that second layer \cite{prbsecondh2}, 0.100 \AA$^{-2}$.  
That density is also of the same order
of magnitude as the corresponding to the uppermost HD layer in a $^3$He/HD/HD, 
studied experimentally in Refs.  \onlinecite{casey2,ikegami1,ikegami2,masutomi,casey1,casey,fukuyama2,fukuyama3}
($\sim$ 0.092 \AA$^{-2}$),  whose results are directly comparable to those of
the present work. 
With all those considerations in mind,  and to avoid as much as possible size effects, we considered a 14 $\times$ 8 first layer cell of 
molecules separated 3.48 \AA$ $ from each other, i.e., a 48.72 $\times$ 48.22 \AA$^2$ simulation cell comprising 224 hydrogen molecules. 
The remaining parameter $b_3$  is 2.96 \AA, as in previous 
literature \cite{preplated} and defines
the Jastrow part of the trial wavefunction, designed to avoid the 
unphysical situation in which two Helium atoms 
are located one on top of each other. 
   
The part of the  trial wave function corresponding to the
layer in contact with the graphite surface,  that contains the different 
hydrogen isotopes is taken as
\begin{eqnarray} \label{trial2}
\psi_1({\bf r}_{1}, \ldots, {\bf r}_{N_1})  = 
\prod_i^{N_1}  u({\bf r}_i)  
\prod_{i<j}^{N_1} \exp \left[-\frac{1}{2}
\left(\frac{b}{r_{ij}} \right)^5 \right]   \times \nonumber  \\
\prod_i^{N_1} \exp \left\{ -a_1 [(x_i-x_{\rm site})^2 + (y_i-y_{\rm site})^2]
\right\} \ .  
\end{eqnarray}
As before, the function $u({\bf r})$ is the numerical solution to the 
Schr\"odinger equation that defines the interaction
between a single hydrogen molecule and the graphite surface, and it depends on the mass of the different species
(H$_2$,HD or D$_2$) on top of the carbon layer. The variational Jastrow  
parameter $b$ was fixed to 3.195 \AA$ $ for all 
isotopes, since in previous literature dealing with similar systems  
\cite{prb2010,prbcarmen,prbsecondh2,prb2022} it was found to be 
independent of the mass.

The last term in Eq. \ref{trial2} pines the atoms around 
their crystallographic
positions (x$_{\rm site}$,y$_{\rm site}$), in this case the ones defining a 
triangular lattice of density 0.095\AA$^{-2}$.  The $a_1$'s parameters 
entering that function depend on the mass of the hydrogen 
adsorbate and to obtain them we
performed several VMC calculations using Eq.  \ref{trial2} as a wavefunction for different values of $a_1$. The 
parameters that produced the lowest energies for the 
different molecular isotopes were $a_1$=1.19 \AA$^{-2}$ (H$_2$), 1.61 \AA$^{-2}$ (HD) and 2.02 \AA$^{-2}$ (D$_2$).  

Eq. \ref{trial3} would define adequately a gas or a liquid  
$^3$He layer. 
On the other hand, to model efficiently a second-layer  $^3$He solid, 
one fixes  those atoms to the crystallographic positions corresponding to the 
structure we are interested in. To do so, we will have to multiply Eq. 
\ref{trial3} by 
\begin{equation}
\prod_i \exp \left\{ -a_2 [ (x_i-x_{\rm site})^2 +  (y_i-y_{\rm
site})^2] \right\} \ ,
\label{trialsol}
\end{equation} 
in which we used the same parameter for all the solid phases and
densities ($a_2 = 0.24$ \AA$^{-2}$)~\cite{preplated,secondhe3}.
We have considered four different phases, three commensurate (the 4/7, widely considered  in the standard literature \cite{casey2,ikegami1,ikegami2,masutomi,casey1,casey,fukuyama2},  the 7/12 \cite{preplated,secondhe3},  and 
the newly proposed 1/2 structure \cite{fukuyama3}) and an 
incommensurate triangular one at different densities. 

The fact that triangular $^3$He solids are incommensurate structures with 
respect 
to those of the first hydrogen layer implies that the dimensions of the 
simulation cells
corresponding to those second-layer solids do not have to be (and in fact, they 
are not) the same as the ones for the hydrogen layer. 
To avoid mismatch problems between those two sheets, we followed the procedure 
reported in Refs. \cite{prbsecondh2,prb2022}. 
First, we used 
as the upper simulation cell the larger piece of a triangular solid of a given
density that fits in the  
48.72 $\times$ 48.22 \AA$^2$ cell defined by the hydrogen substrate.  
For instance, if we consider that the upper density for the $^3$He triangular solid
before the promotion of $^3$He to a second Helium layer is the one given in Ref.  
\onlinecite{fukuyama2}  (0.058 \AA$^{-2}$) the dimensions of that simulation cell are 
44.6 $\times$  46.34 \AA$^2$, corresponding to a 10 $\times$ 6 supercell. To 
take into account all the interactions between any Helium atom and the Hydrogen 
substrate, we replicate the first layer simulation box to create a nine-cell 
structure using the vectors that define that Hydrogen sheet. Then, we 
calculate the corresponding potential terms within a given cutoff distance 
between the particles in the first and second layer, without using 
the minimum image convention. That cutoff must be smaller than half the 
shortest side of the upper simulation cell, in this example 22.3 \AA.
On the other hand, the Helium-Helium interactions are calculated in a similar 
way by using the nine vectors 
(0,0),(0,$\pm$46.34),($\pm$44.6,0),($\pm$44.6,$\pm$46.34) \AA$ $ to replicate 
the initial set of second-layer coordinates using the same 22.3 \AA$ $ cutoff. The 
same recipe was employed for 
the Hydrogen molecules with respect to the second-layer Helium atoms and the 
molecules closer to the graphite surface. 
This procedure makes possible to consider any adsorbate density, and not only
those which fit exactly the periodicity of the first layer. 

In the DMC calculations, $f(\bm{R},t)$ is represented not by an analytical 
function but by a set of \textit{walkers} \cite{boro94,hammond}. Each walker is 
defined by a set of 
coordinates, $\bm{R}$, of all the atoms/molecules of the system. Those positions 
are evolved in imaginary time by the prescription given in Eqs. \ref{eqf} and 
\ref{eqf2} until the local energy of 
the set of particles, $E_L({\bm R})$, varies stochastically around a stable mean 
\cite{boro94,hammond}. That would correspond to the limit $t \rightarrow 
\infty$, limit in which we can calculate other thermodynamic properties. The 
value of those observables is the average over 
the set of walkers. We have checked that considering more than 300 walkers 
leaves the results unchanged. To avoid  
any influence of the initial configurations on the simulations results,  
we typically dispose of the first 
2 10$^4$ Monte Carlo steps (a change in all the particles positions in all the 300 
walkers) 
in a typical 1.2 10$^5$ steps long simulation run. To further avoid spurious influences 
of a particular DMC history, we averaged the energies of three 
independent Monte Carlo runs. 

Finally, to fully characterize the DMC algorithm, we have to bear in mind 
that the $^3$He atoms are fermions. This implies that when we interchange the 
positions 
of any two of those particles with the same spin, the total (and its 
approximation, the trial) function should change sign. We made sure of that by 
introducing the Slater determinants  $D^{\uparrow}$ and 
$D^{\downarrow}$ in Eq. \ref{trial3}. Those determinants impose the nodal 
structure and the positive and negative regions of the wavefunction. In its
simplest form (the one described here and aptly called Fixed-Node diffusion 
Monte Carlo, FN-DMC), 
the algorithm does not change 
the position of those nodes, making the energy derived from it an upper bound
to its exact value \cite{hammond}.  

\section{Results}

\begin{figure}[t]
\begin{center}
\includegraphics[width=0.9\linewidth]{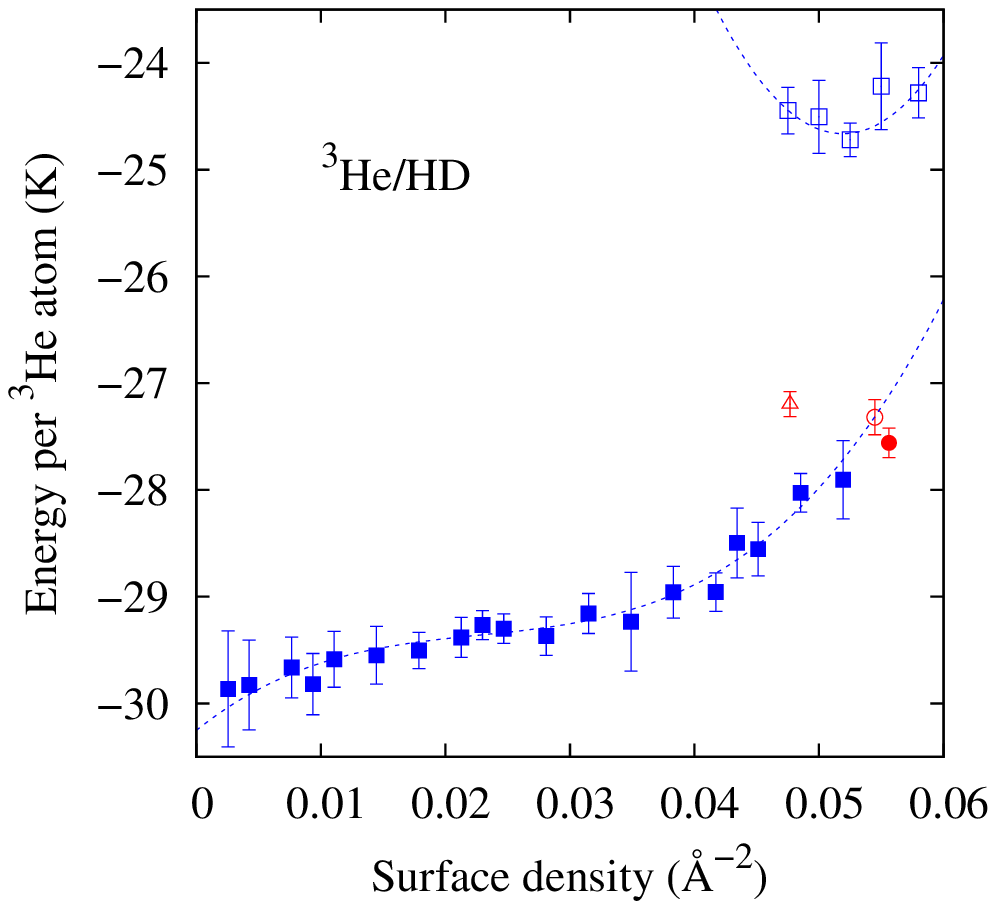}
\caption{Energy per $^3$He atom on a HD layer as a function of the density for
a gas structure (solid squares), for an incommensurate 
triangular solid (open squares), and for different registered phases:
1/2 (open triangle) 4/7 (open circle) and 7/12 (solid circle). The dotted lines 
correspond to least-squares fittings to cubic polynomials and are intended 
mainly as guide-to-the-eye. The density range corresponds to the experimental 
stability region of a single $^3$He absorbed on two HD layers \cite{fukuyama2,fukuyama3}.
}
\label{fig1}
\end{center}
\end{figure}

\begin{figure}[t]
\begin{center}
\includegraphics[width=0.9\linewidth]{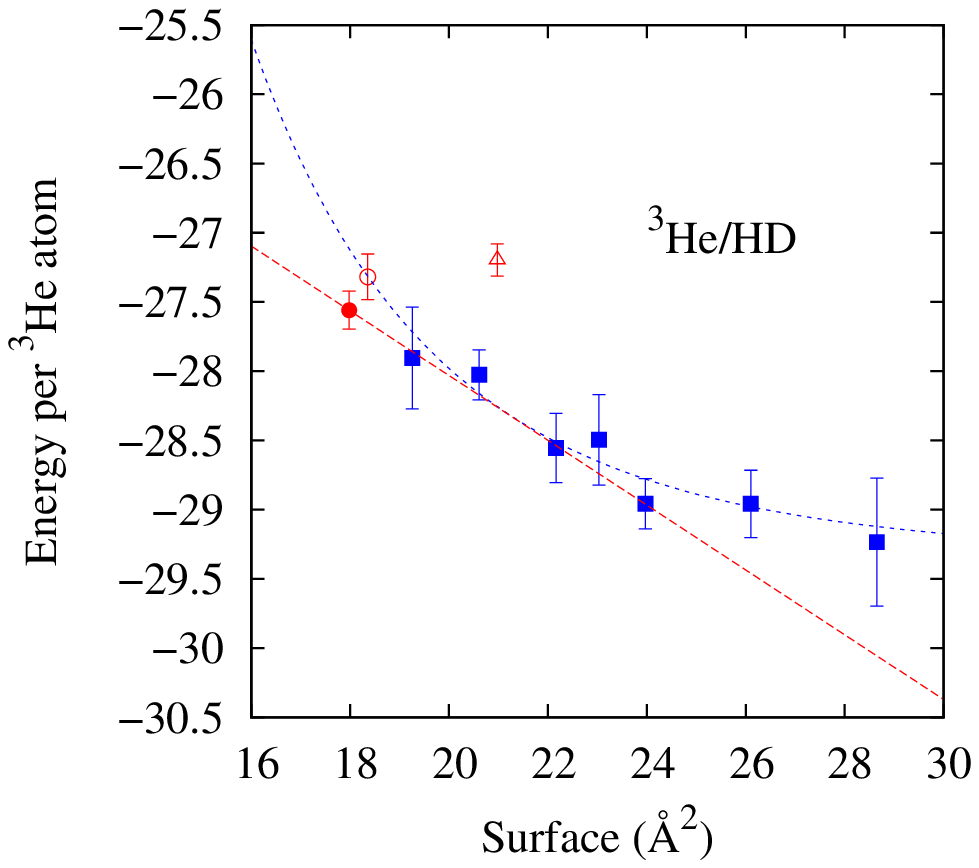}
\caption{Energy per $^3$He atom as a function of the inverse of the density in 
the 
range 0.033-0.060 \AA$^{-2}$. Symbols have the same meaning as in Fig. 
\ref{fig1}. Dashed line is the double-tangent Maxwell construction 
corresponding to 
the equilibrium between a gas of density 0.048 $\pm$ 0.004 \AA$^{-2}$ and a 
7/12 commensurate solid. 
}
\label{fig1b}
\end{center}
\end{figure}

\begin{figure}[t]
\begin{center}
\includegraphics[width=0.9\linewidth]{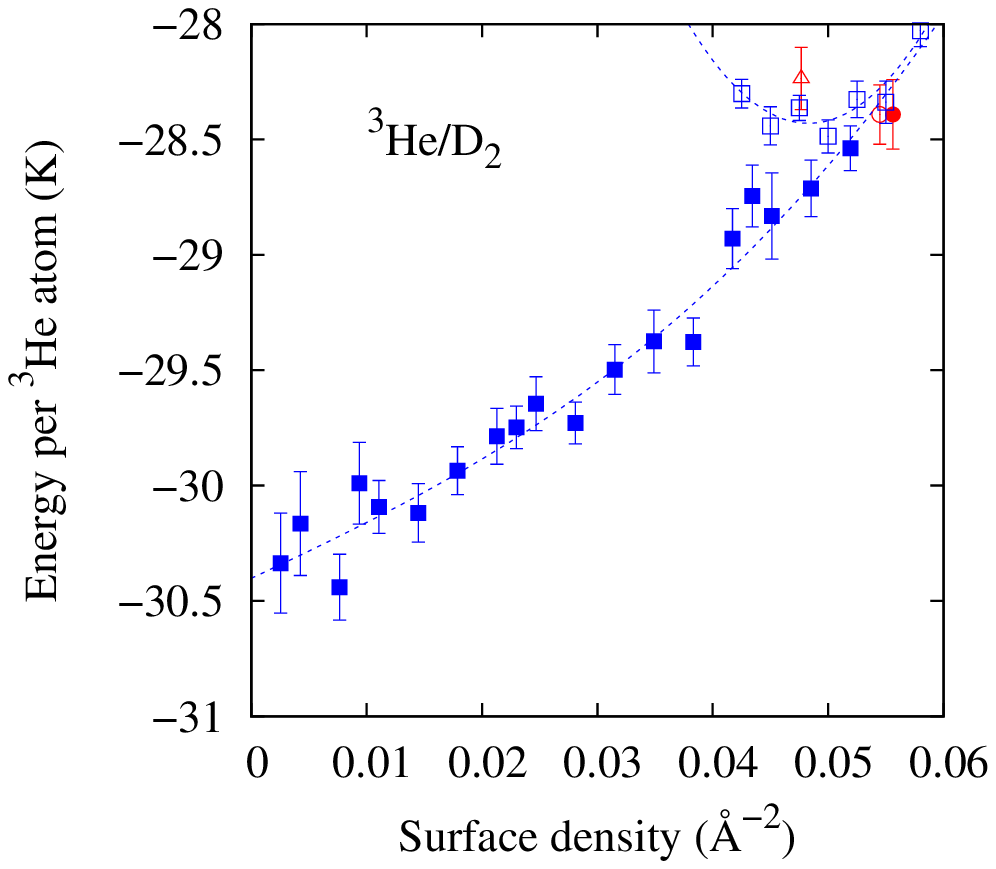}
\caption{Same as in Fig.  \ref{fig1} but for $^3$He on D$_2$.  The symbols and 
lines have the same meaning as in that figure. 
}
\label{fig2}
\end{center}
\end{figure}

\begin{figure}[t]
\begin{center}
\includegraphics[width=0.9\linewidth]{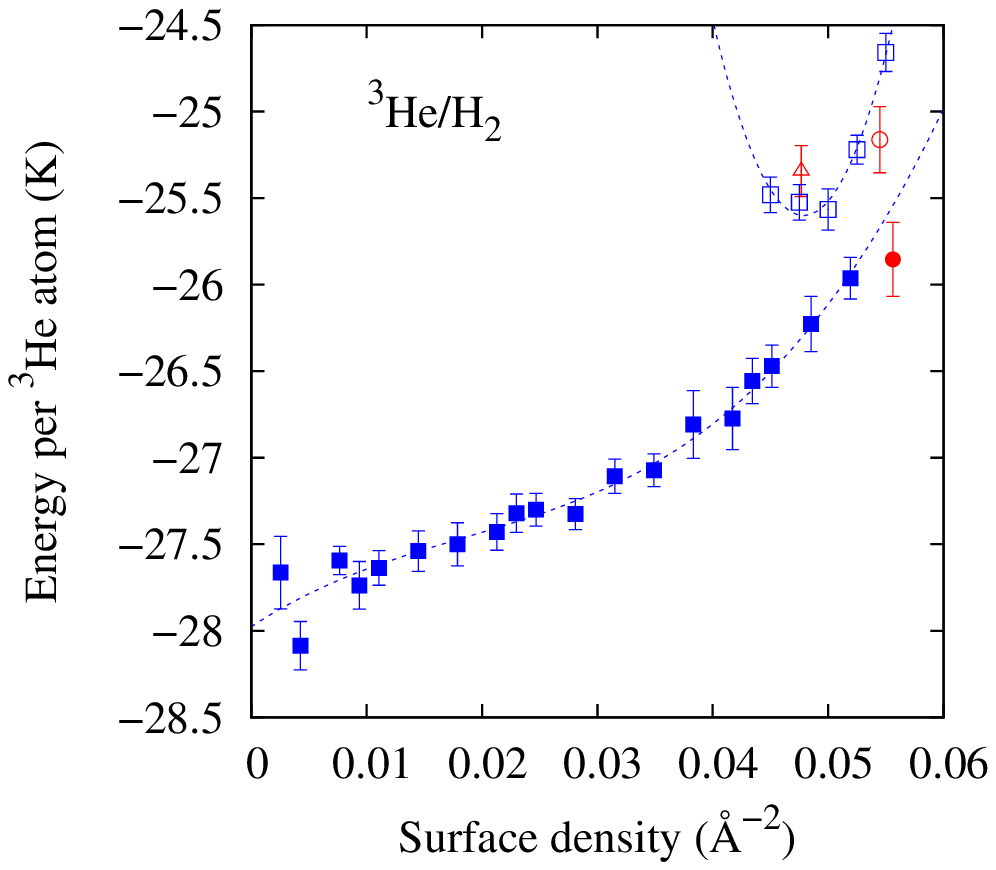}
\caption{Same as in previous figures but for $^3$He on H$_2$.  The symbols and 
lines have the same meaning as in Figs. \ref{fig1} and \ref{fig2}. 
} 
\label{fig3}
\end{center}
\end{figure}

The primary output of any DMC calculation is the energy of the ground state of 
the system under consideration. This implies that we are operating at $T=0$ and 
that the energy is equal to the free energy. From that magnitude, we can obtain 
the 
phase diagram of $^3$He adsorbed on top of the different molecular hydrogen 
substrates. Those results are displayed in Figs. \ref{fig1}-\ref{fig3}. 

Fig. \ref{fig1} displays what happens on the HD surface, the one for which we 
have 
experimental information \cite{casey2,ikegami1,ikegami2,masutomi,casey1,casey,fukuyama2,fukuyama3}. The solid squares correspond to the 
gas phase described by Eq. \ref{trial3} alone, the dotted 
line being the result to a least-squares third-order polynomial fit to that set 
of data. What we observe is that, for that phase, the energy increases 
monotonically as a function of the $^3$He density. Moreover, there is no flat 
region of the curve that we could associate to a liquid-gas transition, as in 
the first layer of $^3$He on graphite \cite{prl2016}, i.e., at low densities the 
stable phase is a gas and not a liquid, at least within the accuracy ($\pm$ 0.3 K) of 
our calculation. This similar to what happens to the second-layer of 
$^3$He on $^3$He on graphite \cite{secondhe3}, but it is at odds with its 
behavior on $^4$He, where a very dilute liquid phase was predicted 
\cite{preplated}.  That diluted phase was labeled as liquid because its energy per 
particle was lower than the corresponding to the infinite dilution limit, i.e., the curve 
of the energy per Helium atom versus density had a local minimum, something not seen here.

The solid structures, described by the product of Eqs. \ref{trial3} and 
Eq. \ref{trialsol}, differ from each other by the set of crystallographic 
positions 
that define them. The energy per Helium atom for a triangular solid is given 
by the open squares in Fig. \ref{fig1}. This phase is clearly unstable with 
respect to both the gas and to any of the other registered phases shown in that 
figure. Those are represented by isolated points: 4/7 (open circle), 7/12 (solid 
circle) and 1/2 (open triangle). This last structure was proposed to be stable 
in Ref. \onlinecite{fukuyama3} and can be built by locating $^3$He atoms on some 
of the potential minima produced by three neighboring Hydrogen molecules 
underneath. In that phase not all such minima are occupied, but only the ones 
needed to produce a honeycomb lattice on the second layer.  Unfortunately, our 
results do not support the stability of that structure,  since its 
energy per atom 
is larger than the corresponding to a gas structure of the 
same density. 

The 4/7 and the 7/12 structures could be stable, though. To check that, 
in Fig. \ref{fig1b}, we display the double-tangent Maxwell construction (dashed 
line) between the 7/12 solid and a gas of density 
0.048 $\pm$ 0.004 \AA$^{-2}$. The slope of that line, that joints the inverse 
density points with the same derivative of the free energy, corresponds to minus 
the equilibrium pressure \cite{chandler}. So, from the two possible Maxwell 
constructions (4/7-gas, not shown,  and 7/12-gas), we have to consider only the 
second, since it corresponds to the lowest pressure value.  This line goes from
20.8 \AA$^2$ (lower surface per particle value for which the gas-like structure is
stable, corresponding to a 0.048 \AA$^{-2}$ helium density) to 18.04 \AA$^2$,  the inverse of the 
density of  the 7/12 registered solid. Taking everything 
into account, we can say that in the 0-0.048 \AA$^{-2}$ range, $^3$He on HD is a 
gas that, upon further Helium loading changes into a 7/12 registered solid of 
density 0.055 \AA$^{-2}$. This is in overall agreement with the experimental 
data in the literature  \cite{casey2,ikegami1,ikegami2,masutomi,casey1,casey,fukuyama2,fukuyama3} and similar to what happens on top of $^3$He 
\cite{secondhe3} and $^4$He \cite{preplated}.  

The DMC algorithm is able to discriminate between $^3$He adsorbed on similar 
substrates, as it can be seen in the comparison between Fig.  \ref{fig1} and 
Figs. \ref{fig2} and \ref{fig3}. 
We can see, for instance, that the Helium binding energies in the respective 
dilution 
limits are different from each other: -30.3 $\pm$ 0.2 K for HD,  -30.4 $\pm$ 0.3 
K for D$_2$ and -28.0 $\pm$ 0.2 K for H$_2$, something that depends exclusively 
on the mass of the molecules of the first layer,  all the interaction potentials 
being equal. This is in agreement to what happens to $^3$He adsorbed on $^4$He \cite{preplated}
and $^3$He \cite{secondhe3},  two substrates with different masses and the same interaction potentials. 
In the first case,   the $^3$He binding energy in the infinitely dilution limit for a 0.112 \AA$^{-2}$ first layer density
was -24.45 $\pm$  0.04 K, to be compared to -22.7 $\pm$ 0.1 K for a layer of density 0.109 \AA$^{-2}$ for the lighter isotope. 
The first value varies very little upon compression of the first layer, increasing to a value of -24.74 $\pm$ 0.07 K 
for an underlying $^4$He density of 0.120 \AA$^{-2}$. This makes us confident that the $\sim$ 2 K difference between 
the binding energy of $^3$He on both helium substrates is due to the mass difference with a very weak dependence on density.  
All of the above implies that,  all things being 
equal, the larger zero-point motion of a first layer lighter isotope produces a smoother effective potential surface in which the local minima a single atom can sit upon are less deep than for more localized isotopes. 
  
\begin{figure}[t]
\begin{center}
\includegraphics[width=0.9\linewidth]{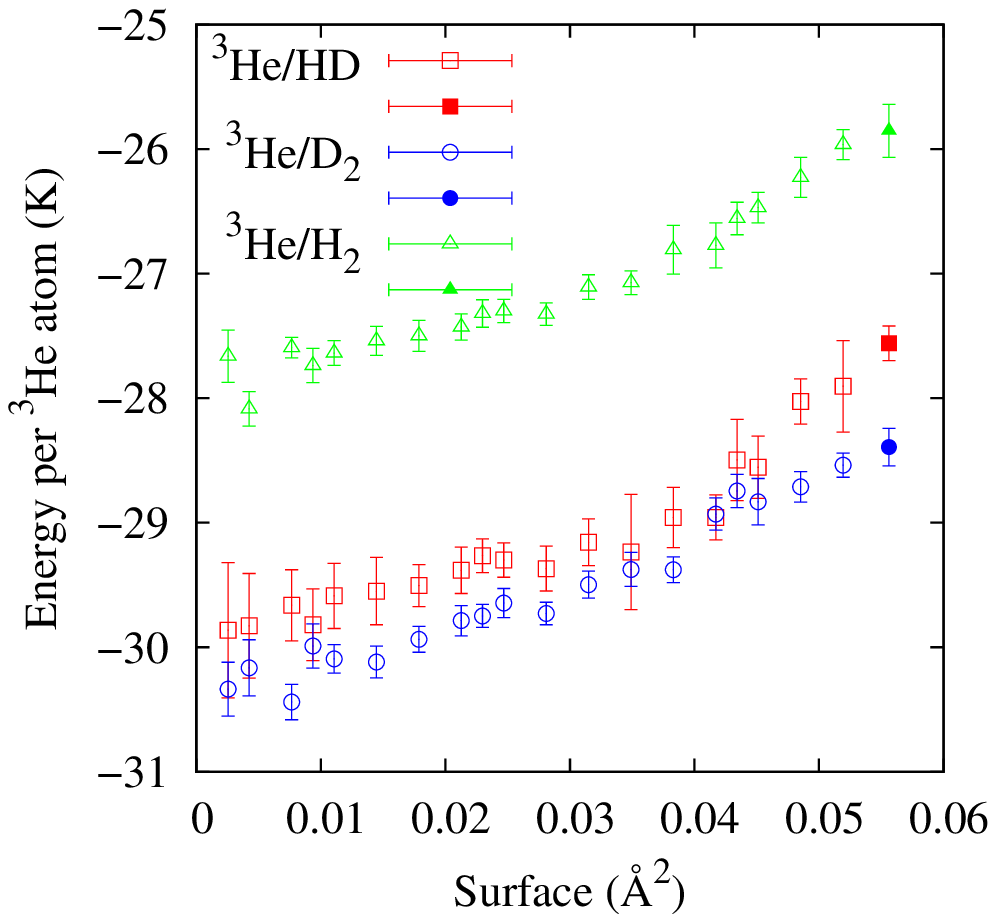}
\caption{Energy per $^3$He atom for the three
substrates considered in this work.  We display only the stable phases to make comparisons easier.  Open symbols,  gas phases; solid symbols,  7/12 structures. 
}
\label{fig4c}
\end{center}
\end{figure}

\begin{figure}[t]
\begin{center}
\includegraphics[width=0.9\linewidth]{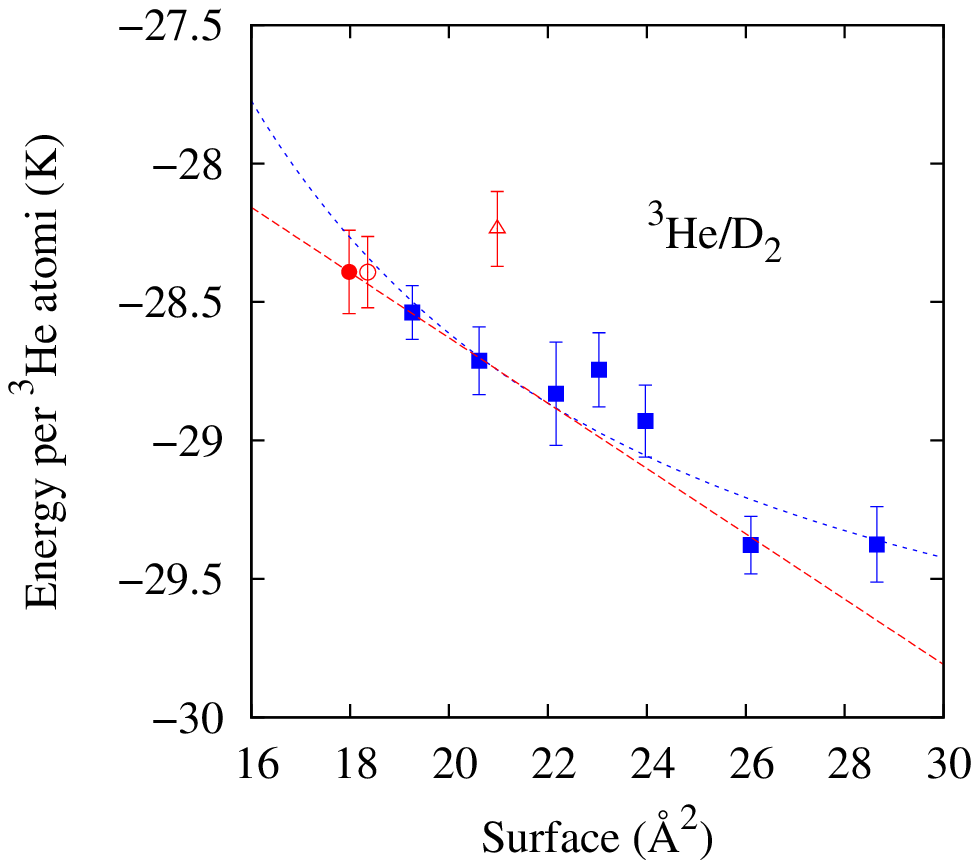}
\caption{Same as in Fig. \ref{fig1b} but for $^3$He on D$_2$
}
\label{maxwelld2}
\end{center}
\end{figure}

\begin{figure}[t]
\begin{center}
\includegraphics[width=0.9\linewidth]{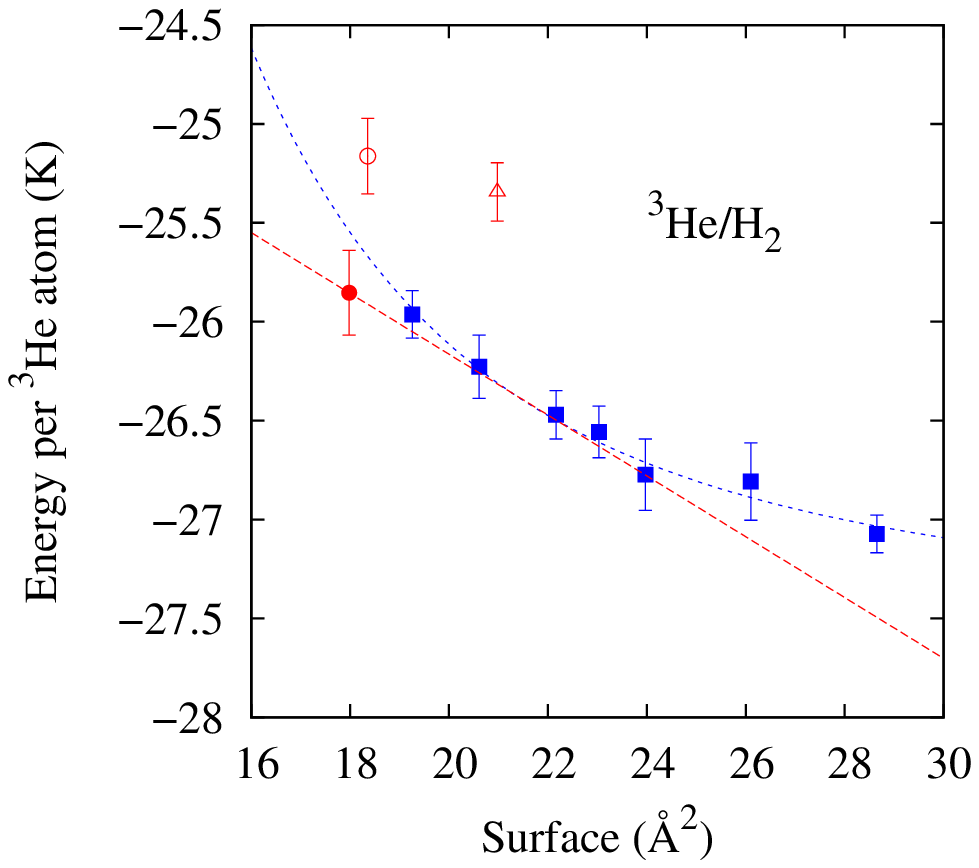}
\caption{Same as in Fig. \ref{fig1b} but for $^3$He on H$_2$
}
\label{maxwellh2}
\end{center}
\end{figure}

The particular details of the dependences of the energy per 
$^3$He atom on the 
second-layer density for the gas phases  
are also substrate dependent,  but the slopes of those curves as a function of the $^3$He density are similar (not equal) to each other, 
as it can be seen in Fig. \ref{fig4c}, in which we display all the stable $^3$He phases for the different hydrogen substrates.  
In addition, by following 
the 
same procedure involving the respective double-tangent Maxwell constructions 
we have found that the stability range for the gas phases is always 0-0.048 
\AA$^{-2}$,  
irrespectively of the Hydrogen isotope (see Figs. \ref{maxwelld2} and \ref{maxwellh2}).  Those gases are also in equilibrium 
with the same 7/12 commensurate phase, all other solid phases being unstable.   The details of the adsorption of the unstable triangular solids are also substrate dependent, but are irrelevant for our conclusions since those phases are not experimentally obtained in the range of densities considered here.  In any case, the fact that the D$_2$ substrate produces a phase whose energy per particle is closest to the gas one can be partially adscribed to the the fact that a non-moving fixed substrate (or one with smaller zero-point displacements) can artificially boost the stability of a solid phase adsorbed on top, as it can be seen in the case of a second layer of $^4$He on graphite \cite{corboz}. 

\section{Conclusions}

By using the DMC method, we were able to calculate the phase diagram of 
$^3$He adsorbed 
on top of a layer of different Hydrogen isotopes. 
In all cases, those diagrams show stable gases in the 0-0.048 \AA$^{-2}$ range in equilibrium with 7/12 registered solids. 
The only difference  would come from the Helium-Hydrogen binding energies, 
something that it can be, but it is hard to be measured 
experimentally.  
In any case, our results compare reasonably well with experimental data for $^3$He/HD/HD, that points to the existence of a  low-density gas phase 
that, upon further Helium loading, changes into a commensurate solid. 
Unfortunately, the nature of that registered phase is different
in this work and in the experiment \cite{fukuyama2}. That difference can be ascribed, at least partially, to the small differences in the densities 
of the underlying first-layer solid and it is comparable to what happens in a second $^3$He layer on helium substrates \cite{preplated,secondhe3}.
On the other hand and importantly, we are able to reproduce the fact that the 
triangular solid is unstable with respect to any other $^3$He phase. 

\begin{acknowledgments}
We acknowledge financial support from Ministerio de Ciencia e Innovación MCIN/AEI/10.13039/501100011033
(Spain) under Grants No. PID2020-113565GB-C22 and No.PID2020-113565GB-C21,
from Junta de Andalucía group PAIDI-205,  and AGAUR-Generalitat de Catalunya 
Grant No. 2021-SGR-01411.  We also acknowledge the use of the C3UPO
computer facilities at the Universidad Pablo de Olavide. We also thank H, Fukuyama by generously share with us the unpublished work on the $^3$He/HD/HD system.
\end{acknowledgments}

\bibliography{hd_boro5}

\end{document}